\documentclass[osajnl,twocolumn,showpacs,superscriptaddress,10pt]{revtex4-1} 
\usepackage{graphicx} 
\usepackage{color} 
\usepackage[normalem]{ulem} 

\begin{document} 
\title{Selective reflection technique as a probe to monitor the growth of a metallic thin film on dielectric surfaces} 

\author{Weliton Soares Martins$^1$, Marcos Ori\'{a}$^1$, Martine Chevrollier} 
\affiliation{Laborat\'{o}rio de Espectroscopia \'{O}tica, DF-CCEN, Cx. Postal 5086 - Universidade Federal da Para\'{i}ba, 58051 - 900 Jo\~{a}o Pessoa - PB, BRAZIL} 
\author{Thierry Passerat de Silans}\email{Corresponding author: thierry@otica.ufpb.br} 
\affiliation{Laborat\'{o}rio de Superf\'{i}cie, DF-CCEN, Cx. Postal 5086 - Universidade Federal da Para\'{i}ba, 58051 - 900 Jo\~{a}o Pessoa - PB, BRAZIL} 

\begin{abstract} 
Controlling thin film formation is technologically challenging. The knowledge of physical properties of the film and of the atoms in the surface vicinity can help improve control over the film growth. We investigate the use of the well-established selective reflection technique to probe the thin film during its growth, simultaneously monitoring the film thickness, the atom-surface van der Waals interaction and the vapor properties in the surface vicinity. 
\end{abstract} 

\maketitle 

\section{Introduction} 
Nanofabrication is a challenge for technology development. On the one hand research has been made in developing lithography with the aim of generating structures in the nanometer range, using, for instance, either electron beam lithography which has reached the few-nanometer range \cite{Craighead83,Chen93} or atomic force microscope to reach ultimate atomic resolution \cite{Bouchiat96}. On the other hand, current-carrying nanowires are used to manipulate cold atoms near a surface and are a promising technique for the development of new devices such as atom chips \cite{Folman00} and for matter waves interferometry \cite{ACHAR}. An alternative approach for such devices is to use the atomic adsorption to construct structures with which the free atoms can interact, for example an array of adatom nanowires to diffract Bose Einstein condensates \cite{McGuirk04}.\\ 

For the optimization of the build up of atomic structures on a surface it is important to get information on the adsorption process \cite{Bordo05} and to control it. For some applications, it might be of interest to control the adsorption in the presence of a vapor at saturated vapor pressure. Resonant light seems to be a good candidate to manipulate the number and position of atoms on the surface \cite{Silans06,Bordo99,Meschede03,Afanasiev07,Burchianti09}.\\ 

To probe a thin metallic film formed at the interface between a dielectric and an atomic gas at vapor pressure, the use of traditional techniques, such as photoelectron spectroscopy, is limited. Moreover, for alkali films, as we are interested here, one cannot easily take away the sample from vacuum environment, because of the reactivity of alkali metals to moisture. It is thus important to develop probing methods compatible with the alkali vapor pressure environment to study \textit{in situ} the film formation \cite{Bordo05}. Light techniques seem to be adequate and, among them, spectroscopic reflection techniques are particularly suitable for studying the interface between transparent dielectric surfaces and alkali vapors. Using reflection techniques with resonant light can also give information on the properties of the vapor in the vicinity of the surface. One approach consists in using the localized field of an evanescent wave (EW) to probe the atomic vapor very close to the surface as well as the thickness of a metallic film on the surface \cite{Bordo05}. Non-resonant techniques such as differential reflectance have also been used to probe film thickness \cite{Proehl05}.\\ 

Selective reflection (SR) spectroscopy is a resonant technique that has been used to probe atomic vapor properties in the vicinity of a surface. The technique was used, for instance, (i) to measure the van der Waals (vW) atom-surface interaction for atomic short-lived excited states \cite{Oria91,Chevrollier92}; (ii) to measure resonant effects between atomic transitions and surface polariton modes that might change the vW interaction into repulsive \cite{Failache99} or induce a surface temperature dependence of the vW force \cite{Gorza06}; (iii) to probe linewidth modifications due to atom-atom collisions \cite{Alkushin83,Sautenkov11}. Furthermore, the SR lineshape is modified if an intermediary layer is introduced between the substrate and the vapor \cite{Vartanyan94,Chevrollier01} and this reflection technique has therefore been suggested as an adequate tool to probe films thickness in such structures. In this article we discuss the appropriateness of the SR-spectroscopy technique to probe, \textit{in situ} and simultaneously, the metallic film thickness, the atom-surface van der Waals interaction and the atomic vapor properties in the surface vicinity.\\ 
  
\section{Selective reflection: theory} 

When a laser beam is sent to an interface between a dielectric surface and a dilute vapor, a change in the reflected intensity occurs when the frequency is scanned across the atomic resonance. The refractive index of the dilute vapor can be written as $n_v=1+\delta n(\omega)$, where $\delta n(\omega)<<1$ is a resonant contribution. The resonant index increment $\delta n=\frac{1}{2}\chi$ is due to the vapor polarization induced by the incident beam, and $\chi$ is the medium effective susceptibility, given at normal incidence by \cite{Ducloy91}: 
\begin{equation} 
\chi=-\frac{2ikN\mu}{\epsilon_0E}\int_0^{\infty}dz\int dv_z  W(v_z)\sigma_{ge}(z,v_z)\exp(2ikz) \label{pz}, 
\end{equation} 

where $\epsilon_0$ is the vacuum permittivity, $k$ is the incident light wavevector, $z$ is the distance from the surface, $E$ is the incident electric field, $N$ is the atomic number density, $\mu$ is the electric dipole moment of the transition and $\sigma_{ge}(z)$ is the coherence term of the density matrix at position $z$. The velocity integration is taken over the Maxwell-Boltzmann distribution, $W(v_z)$, of the velocity component along the beam axis ($v_z$).\\ 

Close to the surface the atomic resonance is shifted by the vW atom-surface interaction which yields thus a position-dependent transition frequency: $\omega_0(z)=\omega_0-C_3/z^3$, where $C_3$ is called the vW coefficient. Pressure-induced homogeneous broadening and frequency shift also change the atomic polarization. Therefore, the observation of the reflected intensity of a laser scanned around the atomic resonance can be used to measure the collisional pressure broadening \cite{Alkushin83,Sautenkov11} and the van der Waals coefficient for excited short-lived levels \cite{Oria91,Chevrollier92,Failache99}.\\ 

A SR spectrum expresses the weighted contribution to the reflected beam of all the atomic velocity classes (see Eq. (1)). In the Doppler limit, where the Doppler linewidth $\Gamma_D$ is much larger than the homogeneous linewidth $\Gamma$ ($\Gamma/\Gamma_D\rightarrow0$), the SR spectrum exhibits thus a logarithmic singularity around zero detuning ($\delta=0$). It has been shown that the use of Frequency Modulation (FM), together with homodyne detection, allows one to obtain a derivative of the SR signal, resulting in a sub-Doppler spectrum with homogeneous linewidth, $\Gamma$ \cite{Alkushin83}. In the Doppler limit, the FM-SR spectra are well described by a single dimensionless parameter $A=\frac{2C_3k^3}{\Gamma}$ \cite{Ducloy91}.\\ 

Considering now a three-layer configuration where a metallic thin film, with complex index $n_2=\alpha+i\beta$, is placed between the dielectric (index $n_1$) and the vapor, the field re-radiated by atomic dipoles in the backward direction is dephased and attenuated compared to the situation without the film. The lineshape of SR spectra is modified and the change in the reflected intensity is given by \cite{Chevrollier01}: 
\begin{equation} 
\Delta R=2R_0\Re\textrm{e}\left(F\delta n\right), 
\end{equation} 
where $R_0$ is the dielectric-vacuum reflectance, and $F(L)$ is an ``attenuation and dephasing reflection coefficient'' that depends on the film thickness (L) \cite{Chevrollier01}: 

\begin{widetext} 
\begin{equation} 
F(L)=\frac{8n_1n_2^2e^{-2in_2kL}}{\left(1+n_2\right)^2\left(n^2_2-n^2_1\right)+2e^{-2in_2kL}\left(1-n^2_2\right)\left(n^2_2+n^2_1\right)+e^{-4in_2kL}\left(1-n_2\right)^2\left(n^2_2-n^2_1\right)}. 
\end{equation} 
\end{widetext} 

The introduction of the complex function $F(L)$ mixes the absorptive ($\Im\textrm{m}(\chi))$ and refractive ($\Re\textrm{e}(\chi)$) atomic responses, distorting the SR spectra as a function of the film thickness.\\ 

\begin{figure}[h] 
	\centering 
		\includegraphics[width=0.90\linewidth]{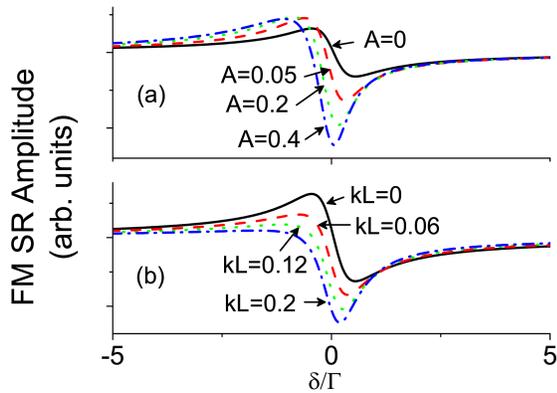} 
	\caption{(Color online) Theoretical FM-SR spectra as a function of the detuning normalized by homogeneous linewidth ($\delta/\Gamma$). The spectra were calculated for (a) different values of parameter $A$ in the absence of metallic film ($L=0$); (b) in the absence of vW interaction (A=0) and for different film thicknesses (values of $kL$).} 
	\label{Fig1} 
\end{figure}

In Fig. \ref{Fig1} we show calculated FM-SR spectra for different values of the dimensionless parameter $A$, without metallic film ($L=0$) (Fig. \ref{Fig1}a) and for different film thicknesses in the absence of vW interaction ($A=0$) (Fig. \ref{Fig1}b).\\ 

The dependence of the SR lineshape on the film thickness, the vdW coefficient and the homogeneous linewidth (typically broadened by pressure effects) may be used as a tool to simultaneously probe, during the film growth, the film thickness, the atom-surface interaction and the vapor in the vicinity of the surface. 

\section{Methodology} 

We discuss in this section the methodology of using SR as a probe of the film thickness and of the vapor properties at the same time. After describing the experimental apparatus, we discuss a method to obtain information through the fitting of experimental spectra by theoretical ones. 

\subsection{Experimental set-up} 
\begin{figure}[h] 
	\centering 
		\includegraphics[width=0.90\linewidth]{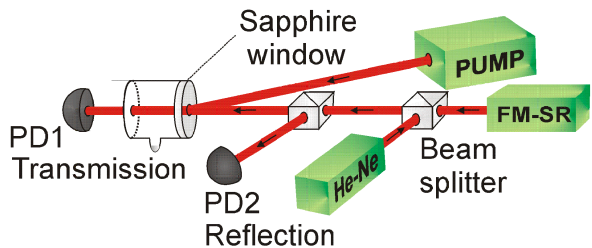} 
	\caption{(Color online) Experimental set-up scheme. The near-resonant pump beam induces the metallic film growth (the angle of incidence is enhanced for visualization). The FM-SR and the He-Ne laser are sent perpendicular to the sapphire window internal surface. The reflected FM-SR beam is detected and sent to the homodyne detection apparatus. The transmitted He-Ne beam is used as a second measurement of the film thickness.} 
	\label{Fig2} 
\end{figure}

The experimental apparatus (see Fig. \ref{Fig2}) is set up to achieve two aims: (i) produce a metallic film on a dielectric surface in a controlled way and (ii) probe the film thickness and the vapor in the vicinity of the surface.\\ 

We want to deposit a Cs metallic film on a sapphire dielectric crystalline surface in an ``atmosphere'' of dilute Cs vapor. For this purpose we have built a T-shaped sealed vacuum-compatible metallic cell with two sapphire viewports connected to the cell body by flanges. The cell was pumped and baked for a week at 300 $^\circ$C in order to avoid residual impurities. After that, a Cs drop was transferred into the cell reservoir and the system was further pumped until it reached the previous pressure. The final residual pressure is less than $10^{-7}$ Torr. The cell is placed in ovens with independent control of temperature for the cell windows and for the cesium reservoir. The cesium reservoir is typically heated to temperatures of $180-220^\circ$C, corresponding to a vapor density of $1-6\times10^{15}$ atoms/cm$^3$. The windows temperature is kept $30-60^\circ$C hotter than the Cs reservoir temperature in order to avoid condensation of cesium on them.\\ 

To deposit the metallic film we use a light-induced lithographic technique \cite{Afanasiev07}, where the film growth is controlled by the intensity of a laser nearly resonant with the atomic vapor. We send to the interface a 5 mm-diameter beam from an amplified diode laser, 900 MHz red-detuned from the cesium D2 line, with power 40 mW, yielding a linear film growth at a rate of 1.3 nm/min. Preliminary spatial filtering of the laser beam gives it a Gaussian shape, ensuring that the film has a smooth thickness variation across the radial axis.\\ 

A second beam (probe laser) from an external-cavity diode laser is frequency-modulated and sent in the normal direction to the dielectric-metallic film-vapor interface. This beam has an intensity of 0.4 mW/cm2 to ensure a linear interaction between the radiation and the vapor. We scan the laser frequency around the Cs $6S_{1/2} (F=4)\rightarrow 6P_{3/2} (F'=3,4,5)$ transitions to perform a FM-SR probing of the film thickness and of the vapor in the surface vicinity. The FM-SR (probe) beam has a diameter of 1 mm, being, therefore, smaller than the film-forming beam, in order to ensure that a film grows with a relatively uniform thickness in the probed area.\\ 

A second independent measurement of the film thickness is provided by a low-power He-Ne laser beam, also sent perpendicular to the interface. Its transmission through the cell is detected by a photodetector. The film thickness is calculated using an expression for the transmission, which considers a metallic thin film placed between two dielectric semi-infinite planes \cite{Tomlin,Dimers}. For these calculations, we have supposed that the optical properties of the Cs film are independent of its thickness. The optical constants values are taken from \cite{Palik}.\\ 

The experiment obeys the following procedure: The film-forming laser is turned on during $5$ min, then it is turned off during 2 min, the duration necessary to proceed with the measurement of the film thickness through both the He-Ne laser transmission and the FM-SR technique. The film is stable during the time where the pumping laser is turned of. The sequence is then repeated. 

\subsection{Fitting the experimental spectra} 
In order to extract information from the SR signals we fit the experimental spectra by theoretical ones. The SR lineshapes are function of many parameters: the vW coefficient ($C_3$), the homogeneous linewidth ($\Gamma$) and the metallic film thickness ($L$). The last two are \textit{a priori} unknown.  The theoretical lineshapes of FM-SR can be calculated as a function of the detuning (normalized by the homogeneous linewidth, $\delta/\Gamma$) for two dimensionless parameters: $A=\frac{2C_3k^3}{\Gamma}$ and $kL$. The multilevel structure of the Cs D2 line is taken into account in our calculations. Three hyperfine excited states are accessible from each hyperfine level of the ground state. Therefore, we sum three identical theoretical curves weighted by their relative transition strength and shifted by the hyperfine splittings. In order to compare a specific theoretical curve ($A$, $kL$) to an experimental spectrum we proceed as follows: (i) we multiply the abscissa axis by the homogeneous linewidth $\Gamma$; (ii) we shift the abscissa axis by a quantity $\Delta$ that corresponds to a frequency shift due to processes other than the vW shift, usually pressure lineshift. The vW shift is implicitly taken into account in the FM-SR lineshape; (iii) we multiply the vertical axis by a constant $C$ that is proportional to $N/\Gamma$ and (iv) we translate the vertical axis by an offset $V$ that is related to the lasers power slope when the frequency is scanned. We use the least-square method to obtain an ensemble of parameters ($\Gamma$, $\delta$, $C$ and $V$) that minimizes the quadratic error between experimental and a theoretical curve. We select the theoretical curve given by the pair of parameters $(A,kL)$ which gives the smaller quadratic error and then obtain from the fitting the desired information about the film and the vapor in the vicinity of the surface: $kL$, $\Gamma$, $\delta$, $N\propto C\Gamma$ and $C_3=\frac{A\Gamma}{2k^3}$.\\ 

\section{Results} 

\subsection{Change in FM SR spectra with film thickness} 
\begin{figure} 
	\centering 
		\includegraphics[width=0.80\linewidth]{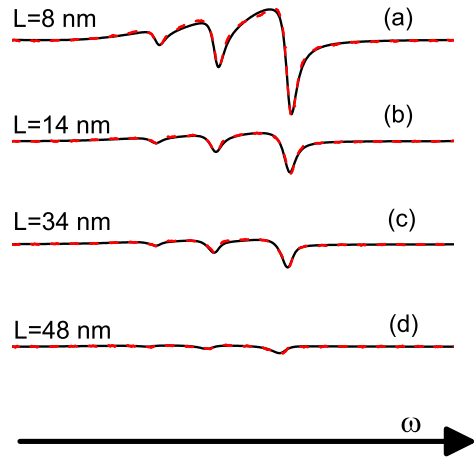} 
	\caption{(Color online) Change of the FM-SR spectra during the metallic thin film growth. Experimental spectra for four different film thicknesses are shown in solid (black) lines. The respective theoretical fits are shown in dashed (red) lines. The film thicknesses are measured using the He-Ne laser transmission.} 
	\label{Fig4} 
\end{figure}

We have acquired FM-SR spectra during a controlled film growth and fitted those curves by theoretical ones (Fig. \ref{Fig4}). The thickness indicated in each figure was obtained measuring the He-Ne laser transmission. At low thickness the FM-SR lineshape essentially has contribution from the real part of $\chi$ and displays a dispersive lineshape (Fig. \ref{Fig4}a). For thicker films, the FM-SR lineshapes become a mixture of absorptive and dispersive contributions \cite{Chevrollier01}, as seen in Fig. \ref{Fig4}(b,c,d). The increase of the film thickness also results in reduction of the signal amplitude, due to the field absorption by the metallic film. The change in the FM-SR spectra shown in Fig. \ref{Fig4} is thus quite pronounced and is expected to allow film-thickness measurements. 

\subsection{$C_3$ and the metallic film thickness} 
\begin{figure} 
	\centering 
		\includegraphics[width=0.95\linewidth]{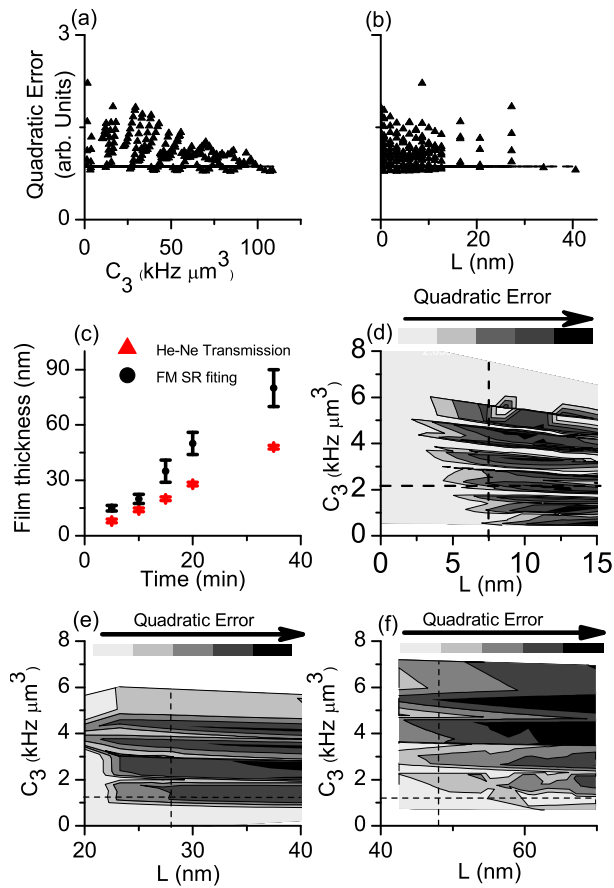} 
	\caption{(Color online) Quadratic error as a function of (a) the vW coefficient $C_3$; (b) the film thickness $L$, which are obtained from the fitting of different theoretical curves to an experimental spectrum. The film thickness for this spectrum was measured with He-Ne beam transmission to be 28 nm.  (c) Evolution of the thickness during a film growth, obtained from fitting FM-SR spectra and from He-Ne transmission. (d) Gray scale plot of the quadratic error as a function of the vW coefficient $C_3$ and of the film thickness obtained from fitting distinct theoretical curves to experimental FM-SR spectra. Those spectra correspond to thicknesses of (d) 8 nm, (e) 28 nm and (f) 48 nm, measured from He-Ne transmission. The dashed horizontal line indicates the theoretical prediction for $C_3$ ($C_3=1.2$ kHz $\mu$m$^3$ \cite{Chevrollier92}).} 
	\label{Fig5} 
\end{figure}

We fit each experimental spectrum by several theoretical curves and analyze the obtained values of $C_3$ and film thickness. In Fig. \ref{Fig5}a we show the quadratic error as a function of $C_3$ obtained from the fitting of an experimental spectrum corresponding to a film thickness of 28 nm (as measured from He-Ne beam transmission). A very large range of $C_3$ values gives the same fitting quality, according to the quadratic error criterion as well as by visual inspection. In Fig. \ref{Fig5}b we plot the quadratic error as a function of the film thickness as extracted from the fitting of the same experimental curve as in Fig. \ref{Fig5}a. There is similarly a large range of L values obtained through fits with the same quality. These results show a relatively large unselectivity on the $C_3$ and film thickness values, which may be attributed to a similar modification of the lineshape for changes of the vW interaction and of the film thickness.\\ 

As we could not obtain simultaneously consistent values of $C_3$ and film thickness, we have fitted a series of experimental spectra imposing a value of $1.0$ kHz$\cdot\mu$m$^3\: < \: C_3\: <\: 2.0$ kHz$\cdot\mu$m$^3$ (the theoretical prediction is $C_3=1.2$ kHz$\cdot\mu$m$^3$ for the Cs D2 line in the sapphire vicinity \cite{Oria91,Chevrollier92}). The thickness values extracted from the fitting of FM-SR spectra as a function of the film formation time are shown in Fig. \ref{Fig5}c, as well as the thickness values deduced from He-Ne laser transmission. We clearly see that the values extracted from the FM-SR spectra are systematically larger than those obtained from He-Ne transmission. To get better insight into the simultaneous measurement of $C_3$ and of film thickness, we plot in gray scale the quadratic error as a function of the mentioned parameters for three experimental spectra: Fig. \ref{Fig5}d for a very thin film of 8 nm; Fig. \ref{Fig5}e and Fig. \ref{Fig5}f for thicker films with thicknesses 28 nm and 48 nm, respectively. We clearly see that the expected $C_3$ and thicknesses values cannot be obtained simultaneously. If one fixes the $C_3$ value to the theoretical one, one gets systematically higher thicknesses values than those measured from He-Ne transmission. For example, in Fig. \ref{Fig5}e the fitting error gets minimum for $L>40$ nm, while He-Ne transmission gives a thickness of 28 nm. A similar conclusion is obtained if one looks for $C_3$ extracted from the fitted curves for the thickness fixed to the He-Ne transmission result, that is, the van de Waals coefficient obtained from the SR spectra fitting are systematicaly larger.\\ 

\subsection{Homogeneous linewidth, collisional shift and density in the surface vicinity} 
\begin{figure} 
	\centering 
		\includegraphics[width=0.8\linewidth]{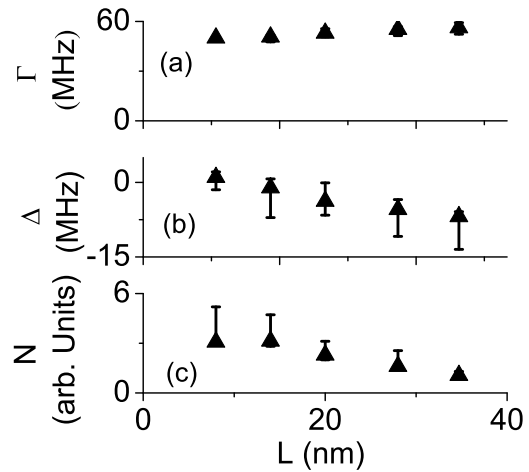} 
	\caption{Changes of (a) the transition homogeneous width $\Gamma$; (b) the pressure shift and (c) the atomic density in the surface vicinity, obtained from fitting theoretical FM-SR curves to experimental spectra, as a function of the film thicknesses obtained from measuring He-Ne transmission.} 
	\label{Fig6} 
\end{figure}

The FM-SR technique gives the homogeneous linewidth and the collisional shift of the atomic transition during the film growth. Besides, the signal amplitude is proportional to the atomic density. Possible changes of the transitions width and shift and of the atomic density with film thickness can thus be detected. In Fig. \ref{Fig6} we show the values of $\Gamma$, $\Delta$ and $N$ obtained from the fitting of experimental spectra as a function of the film thickness (as measured by He-Ne transmission). Figure \ref{Fig6}a shows that the homogeneous linewidth remains almost constant during the film growth. In Fig. \ref{Fig6}b the transition red shift in shown to increase with the film thickness. The atomic density in the surface vicinity decreases as the film grows as shown in Fig. \ref{Fig6}c. These behaviors are not yet clearly understood and need further investigation, but are beyond the scope of the present work. 

\section{Discussion} 

The SR technique was developed to probe the interface between a dielectric surface and a dilute atomic vapor with a smooth metallic film sandwiched between them \cite{Chevrollier01}. Notice, however, that such a model of multiple interfaces is limited because adsorption on ionic surfaces, as for instance, sapphire, does not occur uniformly on the surface \cite{Bonch97,Brause97}. Those surfaces usually exhibit terraces along which the atoms tend to adsorb, forming lines. Moreover, the diffusion of atoms on the surface, including photo-stimulated diffusion, may favor the accumulation of atoms into metallic clusters \cite{Burchianti09}. Thus, the film formed by laser-induced adsorption might be non uniform and possibly clustered \cite{Balzer98,Rasigni73,BonchBruevich00}. The optical properties of the interface are thus different from a dielectric-smooth metallic film - atomic vapor one, with the possibility of near field enhancement and nonlinear atom light interaction. Furthermore, the reflection of the atomic dipole radiation is changed by the presence of the metallic clusters, including possible resonant plasmons effects, which might change the $C_3$ value. The use of such a simple model of the SR technique for such an interface \cite{Chevrollier01} is therefore probably limited.\\ 

In SR experiments at the interface between alkali atoms and a dielectric surface, the experimental values obtained for $C_3$ are usually $1.5-2.0$ larger than the theoretically expected one \cite{Chevrollier92,Failache99,Laliotis07,Segundo}, which considers ideal dielectric surfaces. Indeed, the experiments are done with surface temperatures that are a few tens of degrees above the reservoir temperature in order to avoid alkali condensation on the surface. However, even in these conditions, the surfaces are not ideal dielectrics, and a small quantity of alkali atoms (typically less than a monolayer) is adsorbed on it. Our results indicate that it is necessary to consider this thermal equilibrium coverage of submonolayer metallic film in order to fit the experimental spectra. This may explain the previous systematic disagreement between experimental and theoretical $C_3$ values.\\ 

\section{Conclusion} 
We have investigated the SR technique at a dielectric - metallic thin film - dilute vapor interface, as a tool to probe simultaneously the film thickness, the atom-surface van der Waals interaction and the atomic vapor properties. Our results indicate a very large dispersion in the vW coefficient and film thickness measurements, suggesting that the applicability of this reflection technique demands very careful consideration about the uniformity of the thin metallic film. Our analyze also indicates that the usual disagreement between measured vW coefficient and theoretical expected values may be induced by the presence of adsorbed atoms, and eventually clusters, on the surface. The technique seems to have enough resolution to measure the atomic transitions homogeneous linewidth and lineshift as well as the number of atoms that re-radiate the field. Further quantitative interpretation of experimental SR spectra will rely on a more refined theoretical model taking into account e.g. the discontinuous nature of the film and the variation of the atomic decay rate with the distance to the surface, due to enhanced electric image or non-radiative decay of excitation due to surface quenching. \\ 

{\it Acknowledgements} This work was partially funded by Conselho Nacional de Desenvolvimento Cient\'{i}fico e Tecnol\'{o}gico (CNPq, contract 472353/2009-8), Coordena\c{c}\~{a}o de Aperfei\c{c}oamento de Pessoal de N\'{i}vel Superior (CAPES/Pr\'{o}-equipamentos) and Financiadora de Estudos e Projetos (FINEP). W.S.M and M.C. thank Brazilian agency CNPq for financial support.\\


\begin{thebibliography}{}
\bibitem{Dimers} Notice that part of the He-Ne laser might be absorbed by Cs$_2$ formed by three-atom collisions at those densities. Since the atomic density is not modified during the experiment, the Cs$_2$ absorption does not influence the thickness measurement.
\end{thebibliography}
\end{document}